\documentclass[]{elsart}
\usepackage[dvips]{graphicx}

\begin{document}
\begin{frontmatter}

\title{Advanced simulation code for alpha spectrometry}
\author{T.\ Siiskonen}\footnote{Corresponding author. Tel.\ +358 9 75988 318},
\ead{teemu.siiskonen@stuk.fi}
\author{R.\ P\"oll\"anen}
\address{STUK--Radiation and Nuclear Safety Authority, P.O.\ Box 14, FIN-00881 Helsinki, Finland}
\maketitle

\begin{abstract}
A Monte Carlo code, known as AASI, is developed for simulating energy spectra in alpha spectrometry. The code documented here is a comprehensive package where all the major processes affecting the spectrum are included. A unique feature of the code is its ability to take into account coincidences between the particles emitted from the source. Simulations and measurements highlight the importance of coincidences in high-resolution alpha spectrometry. To show the validity of the simulated results, comparisons with measurements and other simulation codes are presented.
\end{abstract}

\end{frontmatter}

\section{Introduction}

Monte Carlo simulations have proven to be adequate tools for describing alpha-, beta- and gamma-particle transport, even in complex geometries. A great variety of computer codes have been developed for particle transport, dosimetry, particle physics and industrial applications. Different levels of sophistication exist among the codes, but even the simplest ones, which take into account Rutherford and Compton scattering, photoelectric absorption and continuous slowing-down of charged particles, can provide acceptable results. In many cases, simulation is the only practical way to explore the physics behind observed phenomena.

Alpha-particle spectrometry is a widely-used analytical method, for example in surveys of environmental radioactivity. 
The low activity of the samples necessitates long counting times and a small sample-detector distance (SDD). The drawback of a small SDD is the possibility of coincidence summing between the emitted alpha particle and subsequent emissions from the daughter nucleus. In addition, carefully designed sample preparation techniques are essential, since the alpha particles continuously lose their energy as they travel through matter. The energy loss leads to degradation of the spectrum quality via peak spreading, which increases with as the SDD is reduced.

Simulations can be used to investigate the influence of various phenomena on the spectrum quality. The most important factors can be singled out and the measurement setup can be optimised. Moreover, unknown properties of the source, such as source density (or thickness) or source particle properties, can be determined. This is important, especially in the case of direct alpha spectrometry, when radiochemical sample treatment is omitted. The particle beam attenuation and interactions in basic research can also be examined.

Many Monte Carlo simulation packages, such as the TRIM package \cite{bie80}, the GEANT software suite \cite{Geant4}, and MCNP code \cite{mcnp}, are suitable for simulating the alpha particle behaviour in the medium. However, these packages are not necessarily optimal for alpha spectrometry simulations. More specific approaches to alpha-spectroscopic simulations include the backscattering study of Ferrero et al.\ \cite{ferr} and the investigation of aerosol particles by Pickering \cite{pic84}. Rold\'an et al.\ \cite{rol94} examined the spectrum quality at a small SDD. 

The present Monte Carlo simulation code, known as AASI (Advanced Alpha-spectrometric SImulation), is designed to simulate alpha-particle energy spectra. It is intended to be a comprehensive simulation package where all the major processes influencing the energy spectrum are included. Samples of various types (aerosol particles, thick samples, non-uniform samples, etc.) are accommodated. Coincidences between the emitted particles are calculated using nuclide-specific decay data that are stored in a library file prepared in extensible markup language, XML. Although the code has so far been applied to the simulation of alpha particle energy spectra from environmental samples, it can also be used for other applications.
The typical running time on a 1.6 GHz Pentium PC varies from seconds to a couple of minutes depending on the complexity of the simulation problem. The code is written in Fortran 95.

\section{Properties of the source and particle tracking}

Particle propagation through a material layer is determined by two physical processes: direction changes (scattering) and energy loss. The algorithm for particle propagation in a given material layer proceeds as follows:
\begin{enumerate}
\item{Emit a particle from a randomly selected point.}
\item{Calculate the distance, i.e.\ step length to the next scattering (or photoabsorption) event using the cross-section data.}
\item{If the particle is charged, adjust the step if a boundary of absorbing material is crossed. Calculate the continuous energy loss during the step.}
\item{If the particle energy is below the cut-off value, stop tracking.}
\item{If the particle crossed boundary of the material layer, proceed to the next layer if one is present. Otherwise, stop tracking.}
\item{Determine the next direction vector, i.e.\ scattering angles.}
\item{If the particle is a photon, determine the energy loss in the scattering or photoabsorption event.}
\item{Goto (2).}
\end{enumerate}
Here, characteristics of the source as well as the particle tracking method, i.e.\ determination of the scattering angles, are described. Calculation of the energy loss of alpha particles, electrons and photons is presented in the following sections.

\subsection{Source}

Particle emission can originate from a point or from a finite-sized object. These objects, e.g.\ aerosol particles, can be embedded in the source matrix. The composition of the source matrix and the objects that emit radiation need not to be the same. For example, alpha particles can be emitted from an aerosol particle located inside a glass-fibre filter. In the spectrum simulations the number of alpha particle emissions is given in the input.

The thickness of the source can be subjected to random fluctuation that is assumed to follow a Gaussian distribution with a user-given standard deviation $\sigma$. To prevent impossibly large thicknesses, the resulting source thickness $H_\sigma (r)$ is limited to
\begin{equation}
	0 \le H_\sigma (r) \le H + s\sigma,
\end{equation}
where $r$ is the radial position inside the source, $s$ is a user-given parameter and $H$ is the nominal (mean) thickness.

Coordinates of the source particles are sampled as described by Siiskonen and P\"oll\"anen \cite{sii04}. For sources with a random thickness and convex or concave sources, the vertical coordinate is sampled by the rejection method. Convex and concave source shapes are described with a paraboloid of revolution. The user of the code supplies the central and side thicknesses of the source. If the source thickness is zero, all source particles lie on a plane.

Source particles can have a spherical or elliptical shape. Spherical source particles can have a log-normal size distribution. Inactive source particles can be coated with a uniform layer of radioactive material. In addition, a spherical shell of inactive material can be placed around a spherical source particle.

The distance of the source particles from the source surface can be exponentially distributed inside the source matrix. This is a useful feature for investigating air filters in which radioactive aerosol particles are accumulated. This option is only available for cylindrical sources without thickness fluctuations. The distance $d^i$ of a source particle $i$ from the source surface is obtained from
\begin{equation}\label{eq:expdistr}
	d^i = -\lambda\ln\xi^i
\end{equation}
where $\xi^i$ is a random number between 0 and 1, and $\lambda$ is the mean penetration depth given by the user. Another user-given parameter, $f$, determines the fraction of the particles to be distributed according to Eq.\ (\ref{eq:expdistr}). The rest, fraction $1-f$, is distributed on the source surface ($d^i = 0$). Particles which have $d^i$ larger than the source thickness penetrate the source and are ignored. The total number of emissions from the source is reduced accordingly. 

An average solid angle subtended by the detector, the geometrical detection efficiency, is calculated. This is the number of hits received by the detector divided by the number of alpha particle emissions. The desired accuracy, the standard deviation of the efficiency, is given in the input. Calculation of the geometrical detection efficiency is necessary, for example, in direct alpha spectrometry when radiochemical sample treatment is omitted. Tracers cannot then be used for quantitative activity determination.

The measurement setup, consisting of the source, source backing, absorbing material layers and the detector, can be plotted in a file for visual inspection. Library routines for plotting were written by Kohler \cite{koh04}.

\subsection{Particle tracking}

Electrons are tracked when they travel in the source, in the source backing and in the detector, including its dead layer.  Photons are only tracked inside the detector (including its dead layer). Alpha particles are tracked in the source backing for backscattering studies. Otherwise, particles are assumed to travel in straight paths. The tracked particle is followed until it escapes the absorbing material or its energy falls below a cut-off value. When crossing a boundary between two adjacent absorbing layers, the tracking step length is adjusted so that the step does not cross the layer boundary. 

\begin{figure}
\includegraphics[width=\textwidth]{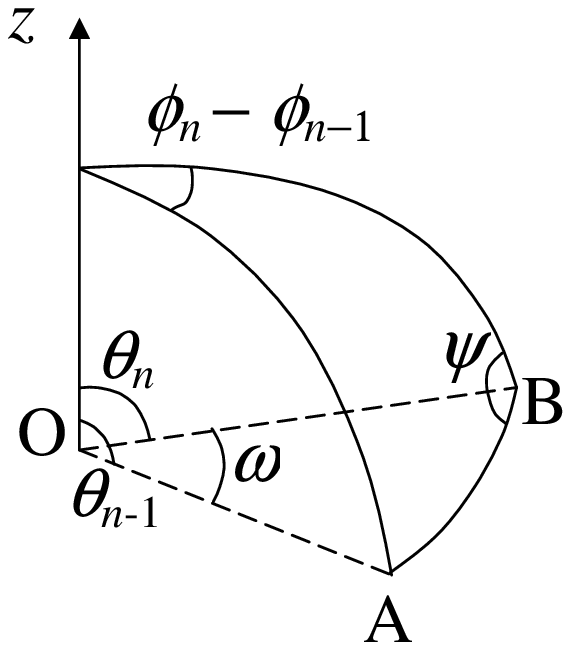}
\caption{Coordinate system. The original particle direction is OA, the new direction is OB. The scattering polar angle is $\omega$ and the azimuthal angle is $\psi$. Laboratory polar angles are $\theta_n$ and $\theta_{n-1}$. The $z$-axis is parallel to the detector symmetry axis and points from the source towards the detector.\label{fig:coords}}
\end{figure}

Particle tracking starts with the sampling of the initial emission coordinates. The initial emission direction $(\theta_0,\phi_0)$ is chosen from a uniform distribution. Following the emission,
the cosine of the polar angle $\theta_n$ (see Fig.\ \ref{fig:coords}) of the tracked particle is determined by
\begin{equation}
	\cos\theta_n = \cos\omega\cos\theta_{n-1} + \sin\omega\sin\theta_{n-1}\cos\psi
\end{equation}
where $\theta_n$ is the polar angle after $n$th scattering, $\omega$ is the scattering polar angle and $\psi$ is the scattering azimuthal angle. The cosine and sine of the azimuthal angle $\phi_n$ are given by
\begin{eqnarray}
	\sin\phi_n &= \sin(\phi_n-\phi_{n-1})\cos\phi_{n-1} + \cos(\phi_n-\phi_{n-1})\sin\phi_{n-1}\\
	\cos\phi_n &= \cos(\phi_n-\phi_{n-1})\cos\phi_{n-1} - \sin(\phi_n-\phi_{n-1})\sin\phi_{n-1}
\end{eqnarray}
where
\begin{equation}
	\sin(\phi_n-\phi_{n-1}) = \frac{\sin\omega\sin\psi}{\sin\theta_n}
\end{equation}
and
\begin{equation}
	\cos(\phi_n-\phi_{n-1}) = \frac{\cos\omega-\cos\theta_n\cos\theta_{n-1}}{\sin\theta_n\sin\theta_{n-1}}.
\end{equation}
All angles are in laboratory coordinates. The scattering angle $\omega$ depends on the differential scattering cross section. After the initial emission, the particles undergo successive scatterings which are assumed to be statistically independent.

Alpha particle scattering is calculated in the centre-of-mass frame. Before the determination of $\cos\theta_n$, the scattering angle is transformed to laboratory coordinates via
\begin{equation}
	\tan\omega = \frac{\sin\varphi}{\cos\varphi + \frac{m_\alpha}{M}}
\end{equation}
where $\varphi$ is the scattering polar angle in centre-of-mass coordinates, $m_\alpha$ and $M$ are the masses of alpha particle and target atom, respectively.

\section{Simulation of alpha particle behaviour in medium\label{sect:alpha}}

\subsection{Energy loss\label{sect:loss}}

Alpha particle energy loss is calculated as described in Ref.\ \cite{sii04}, using the stopping power parametrisation of Ziegler as described in Ref.\ \cite{TOI}. The total stopping power is the sum of the stopping power due to electrons, $S_\mathrm{e}$, and the nuclear stopping power, $S_\mathrm{n}$. In the energy region of interest (below 10 MeV), $S_\mathrm{e}$ is parametrised as
\begin{equation}
	\frac{1}{S_\mathrm{e}} = \frac{1}{S_\mathrm{low}} + \frac{1}{S_\mathrm{high}},
\end{equation}
where
\begin{equation}
	S_\mathrm{low} = c_1E_\mathrm{p}^{c_2} + c_3E_\mathrm{p}^{c_4}
\end{equation}
and
\begin{equation}
	S_\mathrm{high} = c_5E_\mathrm{p}^{-c_6}\ln\left(c_7E_\mathrm{p}^{-1} + c_8E_\mathrm{p}\right).
\end{equation}
Values of parameters $c_1,\dots,c_8$ are tabulated in Ref.\ \cite{TOI}. Here, $E_\mathrm{p}$ is the energy of a proton moving at the same velocity as the alpha particle in question. For composite materials, other parametrisations are also available \cite{sii04}. 

An arbitrary number of absorbing material layers can be added between the source and the detector. The user supplies the number of layers, their atomic and mass numbers, densities, thicknesses and standard deviations representing the thickness fluctuations. Alpha particles are assumed to travel straight paths, except in the source backing where their path is tracked collision by collision. 

\subsection{Energy loss straggling}

Straggling of the alpha particle energy loss can be approximated by a Gaussian energy distribution. Although not strictly correct with thin absorbing layers\footnote{The asymmteric Vavilov (or Landau) distribution is better with thin absorbers. However, computationally they are much more complex.}, it gives a reasonable estimate in many cases. Standard deviation $\sigma_{\rm G}$ of the Gaussian distribution depends on the maximum energy transfer in one collision between the alpha particle and an atomic electron, $E_{\rm max}$, given approximately by $2m_\mathrm{e} c^2\beta^2\gamma^2$. The parameter $\gamma$ is defined through $\gamma m_\alpha c^2 = E$, where $E$ is the energy of the alpha particle. The deviation is given by \cite{Geant4}
\begin{equation}
	\sigma^2_{\rm G} = E_{\rm max}\delta(1-\frac{\beta^2}{2})
\end{equation}
where $\beta$ is the alpha particle velocity in units of $c$. Parameter $\delta$ is the average energy loss in the material layer in question,
\begin{equation}
	\delta = 0.0614\frac{Z}{\beta^2A}\rho\,\delta x\ {\rm keV}.
\end{equation}
Here, $Z$ and $A$ are the atomic and mass number of the target, respectively, $\rho$ is the material density in g/cm$^3$ and $\delta x$ is the distance travelled in micrometers. 

\subsection{Scattering in the source backing plate}

Alpha particles can be tracked in the source backing plate. Screened elastic Rutherford scattering is used to determine the changes in the flight direction. Between the elastic collisions, alpha particles are assumed to lose their energy continuously. The mean free distance between the collisions is calculated from the potential
\begin{equation}\label{eq:rutherfordV}
	V(r) = \frac{e^2}{4\pi\epsilon_0}\frac{2Z}{r}{\rm e}^{-r/a},
\end{equation}
where $a$ is the screening radius, $e$ is electron charge magnitude, $\epsilon_0$ is the permittivity of free space and $r$ is the radial distance. The resulting total cross section is
\begin{equation}\label{eq:rutherfordCS}
	\sigma(E) = \pi \left(\hbar c\alpha\right)^2\frac{Z^2}{E^2\eta(\eta+1)}
\end{equation}
where $\alpha$ is the fine structure constant and screening parameter
\begin{equation}
	\eta =  \left(\frac{\hbar}{a}\right)^2\!\frac{1}{8m_\alpha c^2E}.
\end{equation}
The screening radius is given by \cite{lindhard}
\begin{equation}
	a = \frac{0.885a_0}{\sqrt{z_{\rm eff}^{2/3} + Z^{2/3}}} 
\end{equation}
where $a_0$ is the Bohr radius. The effective charge of the 
alpha particle, $z_{\rm eff}$, is calculated as described in \cite{TOI,ams03}. The mean free distance (step length) between the collisions, $L$, is sampled from
\begin{equation}
	L = -\left(\sigma(E) N\right)^{-1} \ln\xi,
\end{equation}
where $N=N_\mathrm{A}\rho/A$ is the atomic density and $N_\mathrm{A}$ is the Avogadro constant.  The mean atomic spacing $N^{-1/3}$ is used as a step length if $L<N^{-1/3}$. Moreover, if the energy loss between two successive collisions is more than five percent of the alpha particle energy, the step length for the energy loss calculation is reduced until the loss is less than five percent.

Angular deflection in the scattering event from a potential (\ref{eq:rutherfordV}) is given by
\begin{equation}\label{eq:angleSampling}
	\cos\omega = 1 - \frac{2\eta\xi}{1+\eta-\xi}
\end{equation}
and $\psi = 2\pi\xi$ (the $\xi$'s are independent random numbers).

\subsection{Detector response to alpha particles\label{sect:det}}

Alpha particle energy loss in the detector dead layer is treated as described in section \ref{sect:loss}. When the alpha particle hits the active volume of the detector, all its remaining energy is assumed to be deposited. In other words, alpha particles are neither tracked nor is their energy loss calculated in the active volume of the detector. Instead, a simplified solution is chosen which notably reduces the calculation time.

The properties of the detector are read from the user-prepared file. The parameters are the atomic and mass numbers of the detector material, detector radius and thickness, dead layer thickness, detector full-width at half-maximum (FWHM) and the parameters of the exponential tailing function. 

Measurements show that the detector response to monoenergetic alpha particles is not Gaussian \cite{hoir,bortels,stein}. To take this into account, a double-exponential tailing function can be added to the detector response. The resulting energy $E$ is sampled from a distribution
\begin{equation}\label{eq:peakShape}
 P(E) = \frac{\nu R}{1+R}{\rm e}^{-\nu (E_0-E)} + \frac{\mu}{1+R}{\rm e}^{-\mu (E_0-E)},\ E\le E_0
\end{equation}
where $E_0$ is the energy of the incoming alpha particle. The user supplies the parameters $\nu$, $\mu$ and $R$. They should be determined from the measurements of good quality (i.e.\ thin) sources at a large SDD. Parameter $R\ge0$ is the ratio of the areas of the two exponential distributions. Typical values for Canberra PIPS with an area of 450 mm$^2$ are $\nu = 0.1$ keV$^{-1}$, $\mu = 0.02$ keV$^{-1}$ and $R=12.0$. FWHM of the Gaussian detector response is 14 keV. Convolution with the Gaussian detector response is done after the tailing. 

\section{Simulation of electron and photon behaviour in medium}

\subsection{Electrons}

Large-angle deflections of the electrons result from the screened elastic Rutherford scattering, Eqs.\ (\ref{eq:rutherfordV}) and (\ref{eq:rutherfordCS}), with $Z$ replaced with $Z(Z+1)/2$, see, e.g.\ Ref.\ \cite{kij96}. The screening parameter $\eta$ can be chosen from three alternative models. Nigam et al.\ \cite{nig59} suggested that
\begin{equation}\label{nigam}
	\eta_{\rm N} = 5.448\frac{Z^{2/3}}{E_\mathrm{e}},
\end{equation}
where $E_\mathrm{e}$ is the electron kinetic energy. Adesida et al.\ \cite{ade80} fitted electron scattering data in aluminium and proposed that
\begin{equation}
	\eta_{\rm A} = 2.61\frac{Z^{2/3}}{E_\mathrm{e}}.
\end{equation} 
Moli\`ere \cite{mol} (see also \cite{nig59}) concluded that
\begin{equation}
	\eta_{\rm M} = \alpha^2Z^{2/3}\frac{0.36+1.20(\alpha Z/\beta)^2}{\tau(\tau+2)},
\end{equation}
where $\tau = E_\mathrm{e}/(m_\mathrm{e} c^2)$ and $m_\mathrm{e}$ is the mass of the electron.

Between the elastic collisions the electrons continuously lose their energy. The energy loss is calculated by using Bethe's formula \cite{krane}
\begin{eqnarray}
	\frac{dE_\mathrm{e}}{dx} &=& \left(\frac{e^2}{4\pi\epsilon_0}\right)^2\frac{2\pi N_\mathrm{A}Z\rho}{m_\mathrm{e}c^2\beta^2A}\left[
		\ln\frac{E_\mathrm{e}(E_\mathrm{e}+m_\mathrm{e}c^2)^2\beta^2}{2I^2m_\mathrm{e}c^2} \right. \nonumber\\
		&+&\left. 1 - \beta^2 -\ln2\left(2\sqrt{1-\beta^2}-1+\beta^2\right)
		+ \frac{1}{8}(1-\sqrt{1-\beta^2})^2
		\right],
\end{eqnarray}
where $I=16Z^{0.9}$ (eV) is the average ionisation energy of the target atom \cite{Geant4}. We neglect bremsstrahlung and the production of delta electrons and X-rays, since their influence on the detected electron energy spectrum is small. The mean distance between the collisions and the angular deflection is calculated as for the alpha particles, except that the maximum energy loss per step is ten percent of the electron energy.

When an electron hits the detector, its path is followed through the dead layer and into the active volume. In the dead layer, the user has an option to have a partially-depleted region, where the electron deposits part of its energy in the detected signal. The amount of energy deposited in the formation of the signal, $E_{\rm s}$, is then given by
\begin{equation}
	E_{\rm s} = \Delta E_\mathrm{e}\frac{z}{\delta z_{\rm dl}},\ 0\le z\le\delta z_{\rm dl}
\end{equation}
where $\Delta E_\mathrm{e}$ is the electron energy loss at depth $z$ and $\delta z_{\rm dl}$ is the dead layer thickness. In the active volume of the detector, the deposited energy is equal to the detected signal, $E_{\rm s}=\Delta E_\mathrm{e}$.

The number of backscattered and transmitted particles from the detector is calculated. A particle is counted as backscattered when it escapes from the front side of the detector, and as transmitted when it escapes from other sides of the detector. For backscattering studies, a parallel electron beam hitting the detector surface perpendicularly can be used.

\subsection{Photons}

Photons are assumed to interact via photoelectric absorption and Compton scattering. Pair production is ignored, since we are interested in low-energy phenomena. The mean free distance is calculated from the total cross section of the above-mentioned interactions. A random number is used to decide which interaction occurs at the interaction point. 

The total photoelectric absorption cross section is read from a text file and interpolated. Data were obtained from the National Institute of Standards and Technology database \cite{nist}. If data for the element in question do not exist, an analytical approximation for the cross section \cite{scadron}
\begin{equation}
	\sigma_{\rm P}(E_\gamma) = \frac{2^8\pi}{3}\alpha Z^5\left(\frac{E_\mathrm{I}}{E_\gamma} \right)^{7/2}\!a_0^2
\end{equation}
is used. Here, $E_\mathrm{I}=0.0136$ keV and $E_\gamma$ is the photon energy in keV. This approximation overestimates the total cross section for most elements, especially at low energies (less than 100 keV). For example, the overestimation is approximately a factor of 10 in Si when $E_\gamma$ is between 15 keV and 100 keV. As the overestimation is quite large, the user should supplement the photoelectric data library for the element in question, if possible. After the photoelectric absorption, an electron is ejected in the direction of the incoming photon, with $E_\mathrm{e}=E_\gamma$. X-rays produced in this process are ignored. Their energy is often so small that the photon is absorbed in the detector.

The differential cross section for Compton scattering is
\begin{eqnarray}\label{eq:KN}
	\frac{d\sigma_{\rm C}}{d\Omega} &=& \frac{(\alpha\hbar c)^2}{2}\frac{1}{\left[ m_\mathrm{e}c^2+
		E_\gamma(1-\cos\omega) \right]^2} \nonumber \\
		&&\qquad\times\left\lbrace \frac{E_\gamma^2(1-\cos\omega)^2}{m_\mathrm{e}c^2\left[ m_\mathrm{e}c^2+E_\gamma(1-\cos			\omega) \right]} + 1
		+ \cos^2\omega\right\rbrace
\end{eqnarray}
neglecting the binding energy and momentum of the atomic electron. The photon scattering angle is calculated from distribution (\ref{eq:KN}) using the rejection method by Brusa et al.\ \cite{brusa}. The total cross section for Compton scattering is
\begin{eqnarray}
	\sigma_{\rm C}(E_\gamma) &=& \frac{\pi(\alpha\hbar)^2}{m_\mathrm{e}E_\gamma}\left\lbrace\left[1-2\frac{m_\mathrm{e}c^2}{E_			\gamma}
		-2\left(\frac{m_\mathrm{e}c^2}{E_\gamma}\right)^2\right]\ln\frac{2E_\gamma+m_\mathrm{e}c^2}{m_\mathrm{e}c^2}\right.\nonumber \\
		&&\qquad\left. + \frac{1}{2}+4\frac{m_\mathrm{e}c^2}{E_\gamma}-
		\frac{(m_\mathrm{e}c^2)^2}{2(2E_\gamma+m_\mathrm{e}c^2)^2}\right\rbrace.
\end{eqnarray}
The total interaction cross section is $\sigma(E_\gamma) = \sigma_{\rm P}(E_\gamma) + \sigma_{\rm C}(E_\gamma)$. The mean free distance is then
\begin{equation}
	L_\gamma(E_\gamma) = -\left(\sigma(E_\gamma) N\right)^{-1} \ln\xi.
\end{equation}

\section{Treatment of coincidences}

The majority of alpha emitters have a significant decay branch to excited states of the daughter nucleus. The excited states decay by gamma-ray or conversion electron emission. Since the lifetimes of the excited states are typically much shorter than the integration time of the data acquisition electronics, pulse summation between the alpha particle and particles emitted by the daughter nucleus may occur. The summation is more pronounced when the SDD is small. 

The summation may lead to distortion of the peak shape and, thus, may have an influence on nuclide identification. A good example is separation of $^{239}$Pu and $^{240}$Pu, which is difficult even in the case of a high-resolution detector and a sophisticated spectrum deconvolution code. Another example is the coincidence summation of $^{241}$Am, whose main alpha decay branch leads to third excited state of the daughter $^{237}$Np, resulting in a clearly visible bump on the high-energy side of the $^{241}$Am main alpha peak.

The probability of each alpha decay branch is given in the nuclide library file, consisting of a schema file and actual library in XML format. The Fortran-XML interface is written by Markus \cite{Arjen}. The alpha decay branch for an individual decay is selected using a random number. The nuclide library file also contains decay routes of the excited states of the daughter nucleus. Each decay route has a known probability of occurence (yield), decay type (gamma or conversion electron emission), initial and final state indices, and energy of the emitted particle. The emission of a conversion electron is associated with an X-ray, whose energy is given in the library.

For each excited state of the decay route a random number is used to select the next decay channel (i.e., final state and emitted particle). The route is followed until the ground state is reached. For each emitted particle, the emission direction is sampled. After the conversion electron emission, an X-ray is emitted before the cascade is followed further. We assume that each conversion electron is associated with X-ray emission. This is a simplification, since we overlook fluorescence yields and Auger electrons. The approximation is good for heavy elements, whose K-shell fluorescence yields are close to 100\%.

If the particle deposits energy in the active volume of the detector, a coincidence is formed and deposited energy is added to the alpha particle energy. If cascade consists of $n$ subsequent decays, the alpha particle can be in coincidence with $m \le n$ particles. Deposited energies of those $m$ particles are then added to the alpha particle energy. 

The algorithm to calculate the coincidences proceeds as follows:
\begin{enumerate}
\item{Check that decay routes exist, i.e., transitions are available for the present state. If no route is found, exit the loop.}
\item{Use a random number to select the decay route, i.e., decay type, line energy and final state index.}
\item{If the emitted particle is an electron, follow it through the source and its backing. Determine if the particle travels towards the detector.}
\item{If the particle hits the detector, simulate the deposited energy.}
\item{If particle deposits energy to the detector, a coincidence is formed. Add the deposited energy to the alpha particle energy. If the particle is backscattered or transmitted, increase corresponding counters.}
\item{If the emitted particle is an electron, emit the associated X-ray. Go to (3).}
\item{Go to (1).}
\end{enumerate}

The lifetimes of the excited states are available in the library file. However, when coincidences are calculated, they are not taken into account. The lifetimes are assumed to be short enough, compared to the integration time of data acquisition electronics, for a coincidence to be seen.

\section{Comparisons with measurements and other simulations}

\begin{figure}
\includegraphics[angle=270,width=\textwidth]{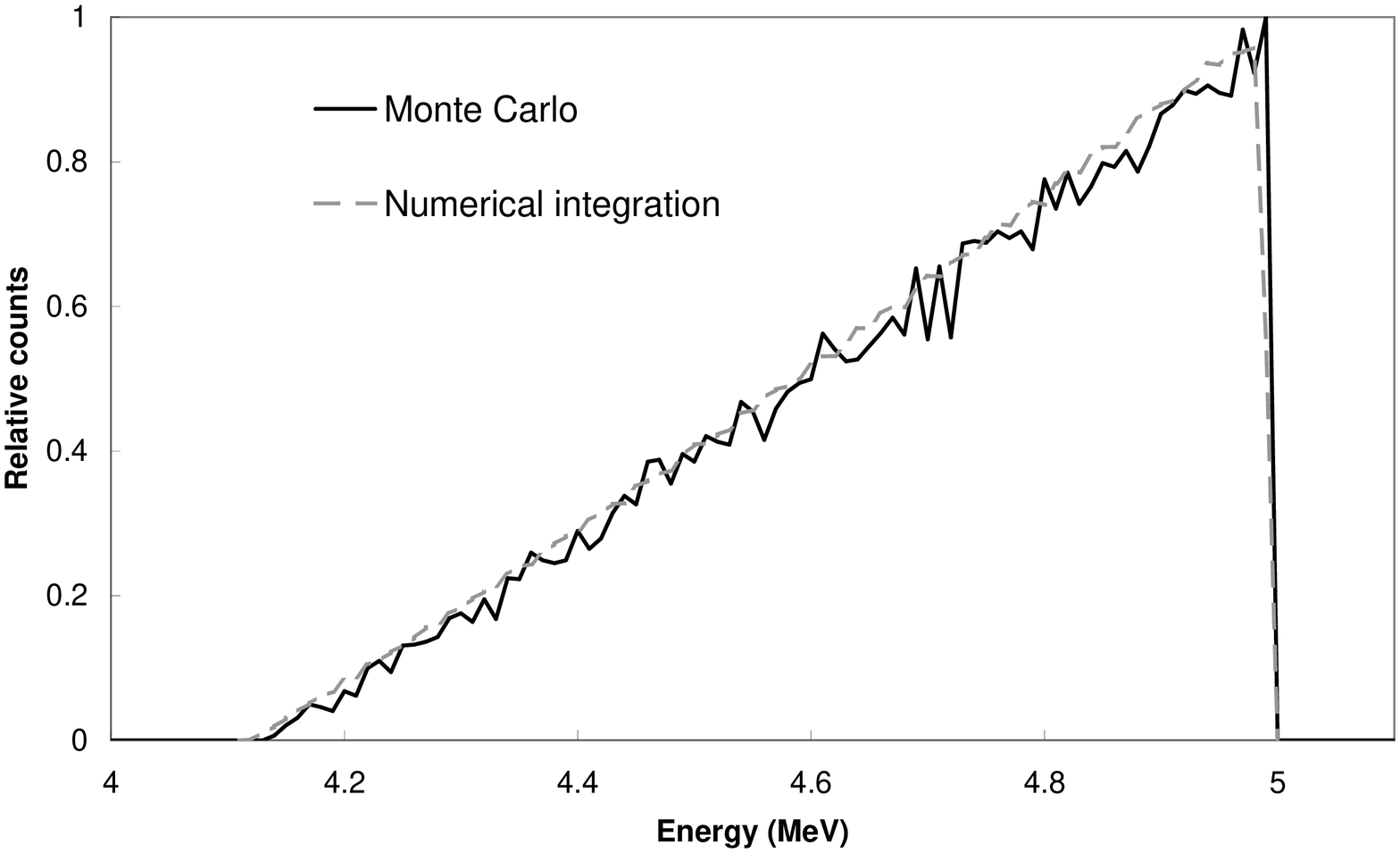}
\caption{Comparison between simulated alpha particle energy spectrum from a convex source (side thickness 0 $\mu$m, central thickness 2 $\mu$m, solid black line) with a spectrum obtained by numerical integration (dashed grey line) from the same source. Alpha particles were assumed to travel in parallel tracks.\label{fig:int}}
\end{figure}

Geometrical detection efficiency and the alpha particle energy loss were investigated by Siiskonen and P\"oll\"anen \cite{sii04}. They found an excellent agreement with earlier results and measurements. To further confirm the homogeneity of the emission point distribution inside a source, we compared a simulated alpha particle energy spectrum from a thick sample with one obtained by numerical integration. In the comparison, a parallel alpha particle beam was considered (corresponding to a very large SDD, polar angle $\theta = 0$) in order to keep the numerical integration tractable. The agreement between the simulated spectrum and the one from numerical integration is good (Fig.\ \ref{fig:int}). This confirms the homogeneity of the emission point sampling, which is qualitatively also shown in Ref.\ \cite{pol05}. Equally good agreement was obtained when the source was assumed to be spherical in shape (results are not shown here).

Electron backscattering can be used to examine the quality of electron transport, since backscattering is sensitively depent on continuous energy loss and angular deflections in elastic Rutherford scattering events. Electron backscattering coefficients for various elements are compared in Table \ref{table:bsc}. The agreement between the experimental values and those of the present work is good. However, the results of the present work are higher than the simulated results of Gueorguiev et al.\ \cite{gue96}. The difference could be explained by different calculation of the continuous energy loss. Gueorguiev et al.\ used different average ionisation energy and a three-point difference scheme. The convergence of the present backscattering calculations was ensured to two significant digits. 

\begin{figure}
\includegraphics[angle=270,width=\textwidth]{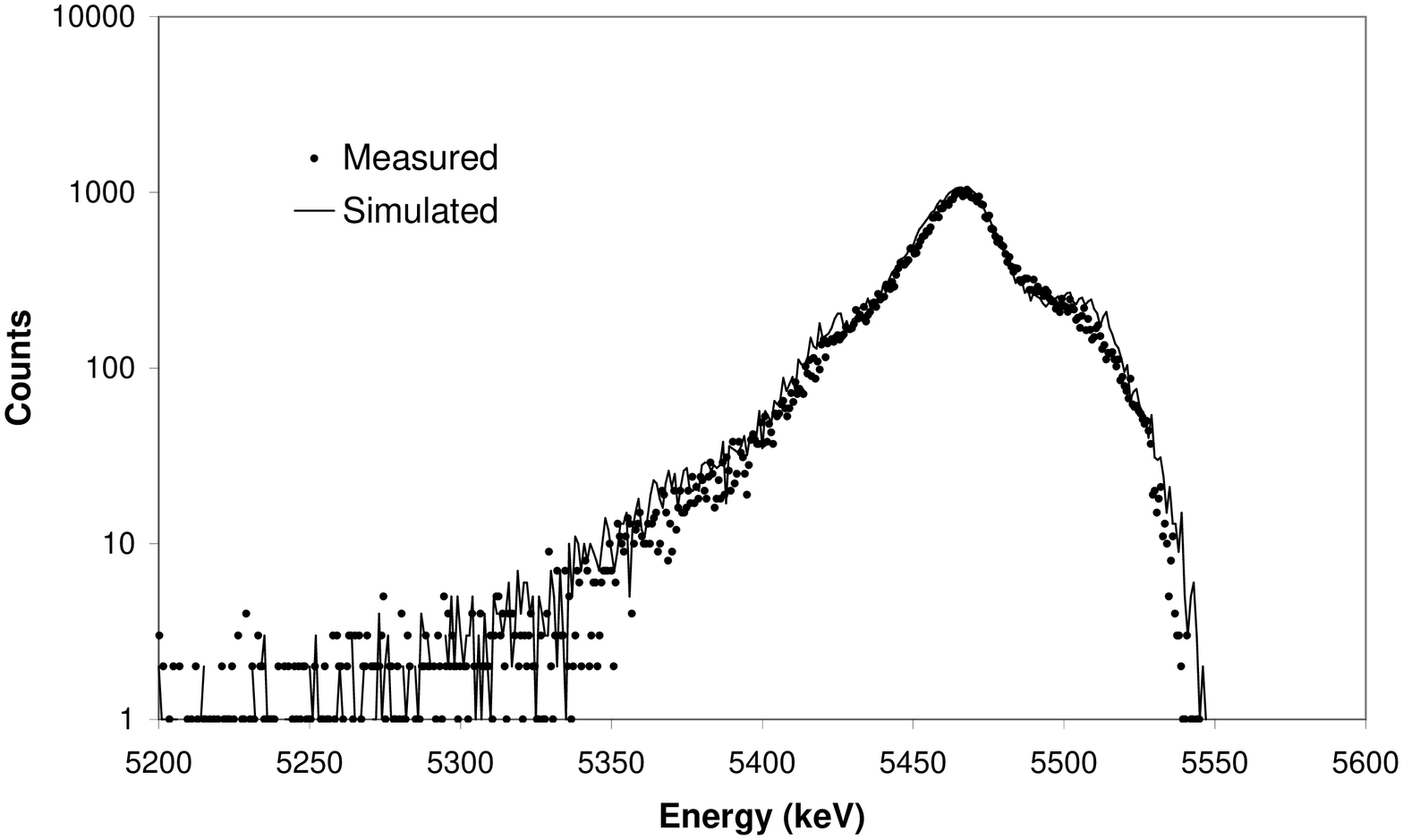}
\caption{Measured (dots) and simulated (solid line) spectrum of $^{241}$Am. The source-detector distance was 5 mm and a detector FWHM of 14 keV was used in the simulation. Detector tailing parameters were $\nu = 0.1$ keV$^{-1}$, $\mu = 0.02$ keV$^{-1}$ and $R=12.0$.\label{fig:am}}
\end{figure}

\begin{figure}
\includegraphics[angle=270,width=\textwidth]{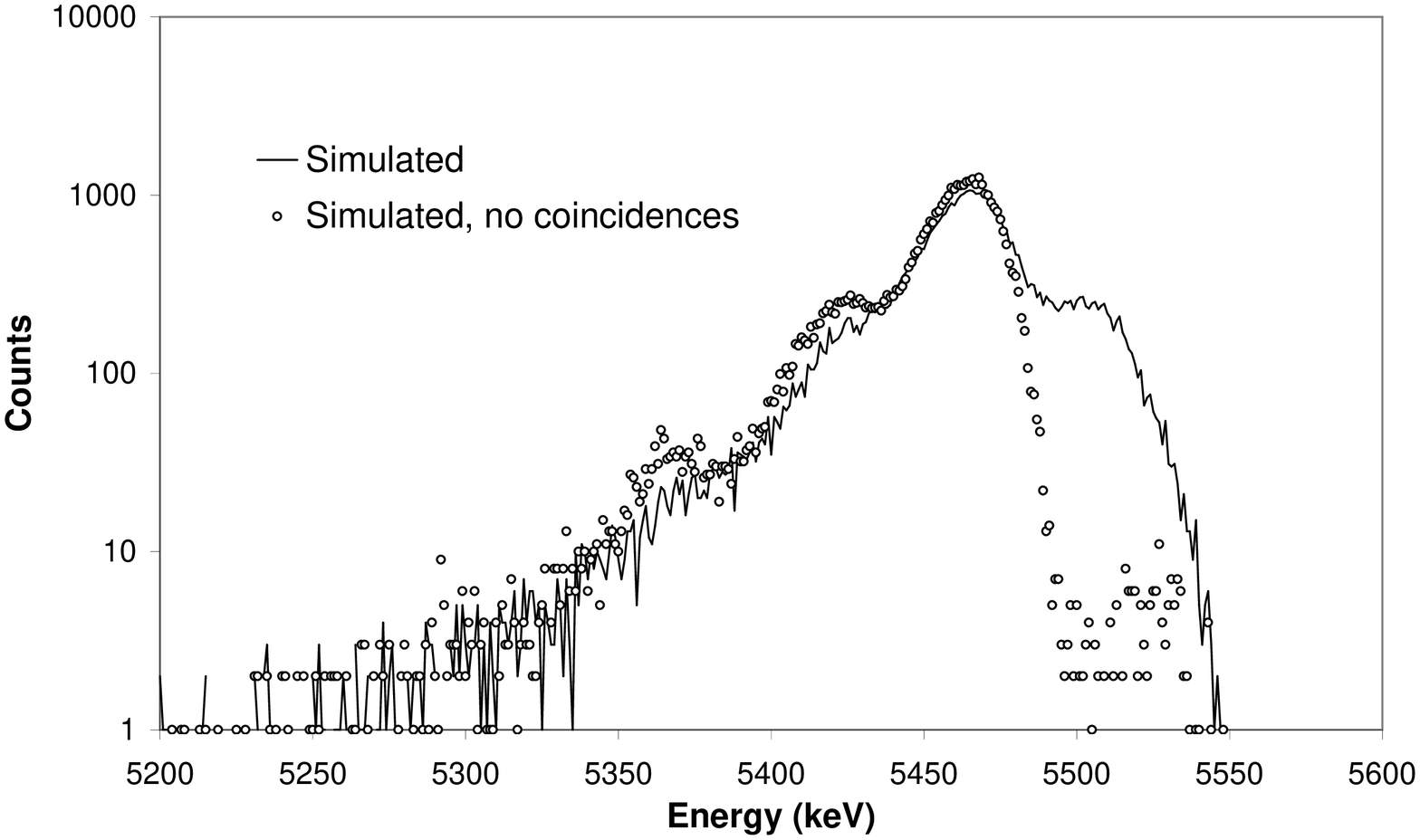}
\caption{Simulated spectra of $^{241}$Am when coincidences are taken into account (solid line, same as in Fig.\ \ref{fig:am}) and when coincidences are ignored (circles). See the caption of Fig.\ \ref{fig:am} for simulation parameters.\label{fig:amNoCoinc}}
\end{figure}

To investigate the influence of coincidences, measured spectrum of $^{241}$Am is compared to the simulation. The coincidence summing of alpha particles with photons and electrons from $^{237}$Np is clearly visible as a bump above 5490 keV (Fig.\ \ref{fig:am}). This bump is absent at a large SDD. Since experimental subshell conversion coefficients for $^{237}$Np were not available, the relative yields for conversion electrons were set as follows: the yield of the L$_\alpha$ line was assumed to be 20\% and that of L$_\beta$ was assumed to be 80\%. Figure \ref{fig:am} illustrates that simulations are accurately able to explain the effect of coincidence summing. When the coincidences are ignored in the simulation, the resulting spectrum clearly disagrees with the measurement (Fig.\ \ref{fig:amNoCoinc}).

\section{Discussion}

The present Monte Carlo simulation code, known as AASI, is designed for simulating energy spectra in alpha spectrometry. The code was originally developed for estimating the influence of source characteristics on the alpha particle spectra, for example in the case when the source quality is not optimal for high-resolution alpha spectrometry. The  sources may be considerably thicker than those obtained from radiochemical sample treatment and their thickness may not be uniform. The source may even be an aerosol filter in which radioactive materials are deposited. This option is useful when the presence of alpha particle emitting materials in the filter must be identified rapidly, i.e.\ alpha particles are counted directly without prior radiochemical sample manipulation \cite{sii04}. This information may be of utmost importance should a nuclear incident or malicious dispersal of radioactive material into the environment occur.

Later development of the code is focused on the development of modelling of the detector and the effect of alpha-electron and alpha-photon coincidences on the measured spectra. The code can be easily used for various alpha detectors provided that the detector response can be treated as the convolution of a Gaussian part of a peak and a double-exponential low-energy tail (Eq.\ \ref{eq:peakShape}). The comparison of measured and simulated spectra, especially in the case of a thin $^{241}$Am source, highlights the importance of coincidence phenomena. When the source-detector distance is small (less than approximately a few cm) the coincidences must be taken into account in unfolding the spectra. Quantitative separation of nuclides such as $^{239}$Pu and $^{240}$Pu in alpha spectrometry is questionable if coincidences are neglected.
Developments in the simulation of electrons and photons in the medium also facilitate the use of AASI in basic research. The detector response to various particle beams can be examined. 

Future development of the code will be addressed in the construction of an appropriate library data file and a user-friendly interface. The validity of the results, i.e.\ the agreement of simulated and those obtained from other calculations or measurements, is verified here and in previous publications. However, validation is a continuous process not limited to this article only. 

An executable version of the code, the input files and a library file needed for the execution of the code are available from the authors upon request.

\clearpage

\begin{table}
\caption{Comparison between experimental (Exp.) and simulated (MC) electron backscattering coefficients for various elements when $E_\mathrm{e}=30$ keV and for normal incidence. In all simulations, the screening model of Nigam et al.\ \cite{nig59} was used, see Eq.\ (\ref{nigam}).\label{table:bsc}}
\begin{center}
\begin{tabular}{lcccc}
\hline
Element & C & Al & Cr & Au \\
\hline
Exp.\ \cite{rei80} & - & 0.144 & - & 0.482 \\
Exp.\ \cite{bis65} & 0.060 & 0.155 & 0.270 & 0.521 \\
MC \cite{gue96} &  0.042 & 0.124 & 0.231 & 0.502 \\
MC, this work & 0.053 & 0.142 & 0.253 & 0.511 \\
\hline
\end{tabular}
\end{center}
\end{table}

\clearpage

\clearpage

\clearpage

\clearpage

\end{document}